\documentclass[twocolumn,aps,superscriptaddress,showpacs]{revtex4}
\usepackage{amssymb}
\usepackage{amsmath}
\usepackage{graphicx}
\usepackage[normalem]{ulem}
\usepackage[dvips]{color}

\def\rpi{$\pi^-/\pi^+$~}

\def\esym{$E_{sym}(\rho)$~}

\renewcommand\sout{\bgroup \color{red} \ULdepth=-.5ex \ULset}

\begin{document}

\title{Circumstantial evidence for a soft nuclear symmetry energy at supra-saturation densities}

\author{Zhigang Xiao}
\affiliation{Department of Physics, Tsinghua University, Beijing
100084, P.R. China}
\author{Bao-An Li\footnote{Corresponding author,
Bao-An\_Li$@$Tamu-Commerce.edu}} \affiliation{Department of
Physics, Texas A\&M University-Commerce, Commerce, Texas
75429-3011, USA}
\author{Lie-Wen Chen}
\affiliation{Institute of Theoretical Physics, Shanghai Jiao Tong
University, Shanghai 200240, P.R. China}
\author{Gao-Chan Yong}
\affiliation{Institute of Modern Physics, Chinese Academy of
Sciences,  Lanzhou 730000, P.R. China}
\author{Ming Zhang}
\affiliation{Department of Physics, Tsinghua University, Beijing
100084, P.R. China}

\date{\today}

\begin{abstract}
Within an isospin- and momentum-dependent hadronic transport model
it is shown that the recent FOPI data on the $\pi^-/\pi^+$ ratio in
central heavy-ion collisions at SIS/GSI energies (Willy Reisdorf
{\it et al.}, NPA {\bf 781}, 459 (2007)) provide circumstantial evidence
suggesting a rather soft nuclear symmetry energy \esym at $\rho\geq 2\rho_0$ compared to the
Akmal-Pandharipande-Ravenhall prediction. Some astrophysical
implications and the need for further experimental confirmations are
discussed.
\end{abstract}

\maketitle

The density dependence of the nuclear symmetry energy \esym is
critical for understanding not only the structure of rare
isotopes\cite{Bro00}and heavy-ion
reactions\cite{LiBA98,LiBA01b,Dan02a,Bar05,LCK08}, but also many
interesting issues in astrophysics\cite{LCK08,Sum94,Lat04,Ste05a}.
To determine the \esym and thus the equation of state (EOS) of
neutron-rich nuclear matter has been a longstanding goal of both
nuclear physics and astrophysics. While both fields have some
promising tools for probing the \esym over a broad density range,
they all have some limitations. Thus, only by combining carefully
complementary information from both fields will we have ultimately
a good understanding about the \esym. While significant progress
has been made over the last few years in constraining the \esym at
sub-saturation densities using terrestrial nuclear laboratory
data\cite{LCK08}, still very little is known about the \esym at
supra-saturation densities.

In fact, the high density behavior of the \esym has long been
regarded as the most uncertain property of dense neutron-rich
nuclear matter\cite{Kut94,Kub99}. Presently, at supra-saturation
densities even the trend of the \esym, i.e., whether it increases
continuously or decreases at some point with the increasing density
is still controversial. While many microscopic and/or
phenomenological many-body theories using various interactions, such
as the Relativistic Mean Field~\cite{Che07} and
Brueckner-Hartree-Fock approaches~\cite{LiZH06}, predict that the
\esym increases continuously at all densities, an approximately
equal number of other models including the Variational Many-Body
theory~\cite{Pan72,Fri81,Wir88a} and the
Dirac-Brueckner-Hartree-Fock\cite{Kra06} predict that the \esym
first increases to a maximum and then may start decreasing at
certain supra-saturation densities depending on the interactions
used~\cite{Szm06}. Additionally, the non-relativistic Hartree-Fock
(HF) approach using many Skyrme~\cite{Bro00,Cha97,Sto03,Che05b},
Gogny~\cite{Dec80}, Myers-Swiatecki~\cite{MS} and the
density-dependent M3Y interactions~\cite{Kho96,Bas07} or the
non-relativistic Thomas-Fermi approach using the Seyler-Blanchard
interaction~\cite{Ban00} also lead to decreasing \esym starting at
some supra-saturation densities. Thus, currently the theoretical
predictions on the \esym at supra-saturation densities are extremely
diverse. Therefore, to make further progress in determining the
\esym at supra-saturation densities, what is most critically needed
is some guidance from experiments. In this Letter, we report
circumstantial evidence suggesting a rather soft
nuclear symmetry energy \esym at $\rho\geq 2\rho_0$ compared to the
Akmal-Pandharipande-Ravenhall (APR) prediction\cite{Akm98} based on a transport model
(IBUU04\cite{IBUU04}) analysis of the recent FOPI data on the
$\pi^-/\pi^+$ ratio from central heavy-ion collisions at SIS/GSI energies\cite{Rei07}.

\begin{figure}[h]
\vspace{-0.3cm}
\centering\includegraphics[width=0.8\columnwidth,scale=0.6]{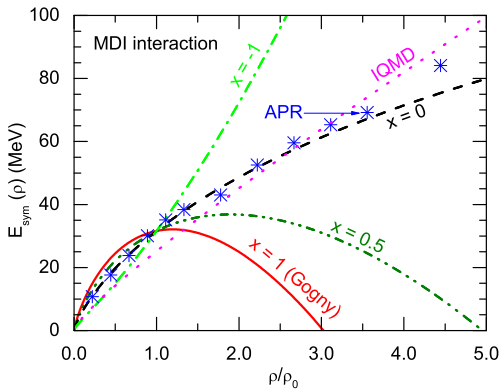}
\vspace{-0.4cm} \caption{(Color online)Density dependence of nuclear
symmetry energy \esym predicted by APR (stars),
used in IQMD (dotted line) and that using the MDI interaction with x=1, 0, 0.5 and -1, respectively.}
\vspace{-0.3cm} \label{fig:esym}
\end{figure}

The isospin and momentum-dependent mean field potential (MDI) used
in the IBUU04 reads\cite{Das03}
\begin{eqnarray}
U(\rho, \delta, \textbf{p},\tau)
=A_u(x)\frac{\rho_{\tau^\prime}}{\rho_0}+A_l(x)\frac{\rho_{\tau}}{\rho_0}\nonumber\\
+B\left(\frac{\rho}{\rho_0}\right)^\sigma\left(1-x\delta^2\right)\nonumber
-8x\tau\frac{B}{\sigma+1}\frac{\rho^{\sigma-1}}{\rho_0^\sigma}\delta\rho_{\tau^{\prime}}\nonumber\\
+\sum_{t=\tau,\tau^{\prime}}\frac{2C_{\tau,t}}{\rho_0}\int{d^3\textbf{p}^{\prime}\frac{f_{t}(\textbf{r},
\textbf{p}^{\prime})}{1+\left(\textbf{p}-
\textbf{p}^{\prime}\right)^2/\Lambda^2}}
\end{eqnarray}
where $\rho_n$ and $\rho_p$ denote the neutron ($\tau=1/2$) and
proton ($\tau=-1/2$) density, respectively, and
$\delta=(\rho_n-\rho_p)/(\rho_n+\rho_p)$ is the isospin asymmetry of
the nuclear medium. All parameters in the above equation can
be found in refs.\cite{IBUU04}. The variable $x$ is introduced to
mimic different forms of the \esym predicted by various many-body
theories without changing any property of the symmetric nuclear
matter and the \esym at normal density $\rho_0$. The \esym with $x$
values of $1, 0.5, 0$ and $-1$ are shown in Figure \ref{fig:esym}.
By setting $x=1$ one recovers the HF prediction using the original
Gogny force\cite{Das03}. As it is well known, the latter predicts a
soft \esym decreasing with increasing density at supra-saturation
densities. For comparisons, shown also are the
$E_{sym}(\rho)=12.5\rho/\rho_0+12.7(\rho/\rho_0)^{2/3}$ used in the
IQMD (Isospin-Dependent Quantum Molecular Dynamics)\cite{IQMD} and
the well-known APR prediction\cite{Akm98}. The latter has been widely used in
calibrating other model calculations. In terms of reproducing the
experimental data, the IBUU04 model has had modest successes so
far\cite{LCK08}. While the NSCL/MSU isospin diffusion
data\cite{Tsa04} allowed us to limit the \esym at sub-saturation
densities to be between that with $x=0$ and
$x=-1$\cite{Chen05,Liba05}, the same model parameter sets
underpredict~\cite{Liba06} significantly the double
neutron/proton ratio of ref.\cite{Fam06}. Nevertheless, it is
interesting to mention that the above limited range of the \esym for
$\rho\leq \rho_0$ is consistent with that extracted very recently
from analyses using the ImQMD (Improved QMD) model which can
reproduce both the isospin diffusion data and the double
neutron/proton ratio simultaneously\cite{Tsa08}. It is also worth
noting that the APR prediction for the \esym at sub-saturation
densities lies right between that with $x=0$ and $x=-1$.

Among the most sensitive probes of the \esym at supra-saturation
densities proposed in the literature\cite{LCK08}, the $\pi^-/\pi^+$
ratio in heavy-ion collisions is particularly promising.
Qualitatively, the advantage of using the $\pi^-/\pi^+$ ratio is
evident within both the $\Delta (1232)$ resonance model\cite{Sto86}
and the statistical model\cite{Ber80} for pion production. Assuming
only first chance inelastic nucleon-nucleon collisions produce pions
and neglecting their re-absorptions, the $\Delta$ resonance model
predicts a primordial $\pi^-/\pi^+$ ratio of $(\pi^-/\pi^+)_{\rm
res}\equiv (5N^2+NZ)/(5Z^2+NZ)\approx (N/Z)^2_{\rm dense}$, where
the $N$ and $Z$ are neutron and proton numbers in the participant
region of the reaction. The $\pi^-/\pi^+$ ratio is thus a direct
measure of the isospin asymmetry $(N/Z)_{\rm dense}$ of the dense
matter formed. The latter is determined by the \esym through the
dynamical isospin fractionation\cite{LiBA05a}, namely, the high(low)
density region is more neutron-rich (poor) with a lower \esym at
supra-saturation densities. Since effects of the \esym are obtained
mainly through the corresponding nuclear mean-field which dominates
the dynamics of heavy-ion reactions at relatively low energies,
based on the resonance model one thus expects the $\pi^-/\pi^+$
probe to be most effective at beam energies near the pion production
threshold $E^{\pi}_{th}\approx 300$ MeV.
On the other hand, assuming pions have gone through multiple
production-reabsorption cycles and reached thermal-chemical
equilibrium, the statistical model predicts that $\pi^-/\pi^+\propto
{\rm exp}\left[2(\mu_n-\mu_p)/T\right]={\rm exp}\left[8\delta
E_{sym}(\rho)/T\right]$, where $T$ is the temperature. Thus, in this
model the $\pi^-/\pi^+$ ratio measures directly the \esym at the
pion freeze-out. Meanwhile, at energies much higher
than the $E^{\pi}_{th}$ where pions are abundant, the reaction
dynamics is dominated by scatterings among all hadrons instead of
the nuclear mean-field\cite{LiBA91}. Therefore, one expects that the
$\pi^-/\pi^+$ probe becomes less effective at very high energies
where other observables, such as, the neutron-proton differential
flow\cite{LiBA00,LiBA02}, may be more useful for probing the high
density \esym. More quantitatively and realistically compared to the
above two idealized models, several hadronic transport models have
shown consistently that the $\pi^-/\pi^+$ ratio is indeed sensitive
to the \esym\cite{LiBA02,Gai04,LiQF05b} especially near the
$E^{\pi}_{th}$. Moreover, by varying separately the \esym at sub-
and supra-saturation densities in IBUU04 simulations we found
that the $\pi^-/\pi^+$ ratio in collisions near the $E^{\pi}_{th}$
is much more sensitive to the variation of the \esym at supra-saturation rather than
sub-saturation densities.

Recently, Reisdorf et al.\ studied systematically the
$\pi^-/\pi^+$ ratio in $^{40}$Ca+$^{40}$Ca, $^{96}$Ru+$^{96}$Ru,
$^{96}$Zr+$^{96}$Zr and $^{197}$Au+$^{197}$Au reactions using the
FOPI detector at SIS/GSI\cite{Rei07}. Their $\pi^-/\pi^+$ data are
the most extensive and accurate ones available in the literature,
thus providing us the best opportunity so far to extract the \esym
at supra-saturation densities.

\begin{figure}[h]
\vspace{-0.2 cm}
\centering\includegraphics[width=0.8\columnwidth,scale=0.6]{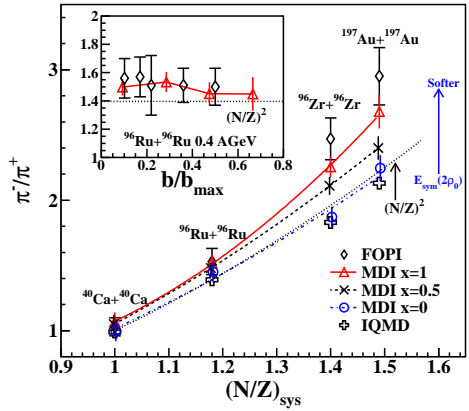}
\vspace{-0.4cm} \caption{(Color online)The $\pi^-/\pi^+$ ratio as a
function of the neutron/proton ratio of the reaction system at 0.4
AGeV with the reduced impact parameter of $b/b_{max}\leq 0.15$. The
inset is the impact parameter dependence of the $\pi^-/\pi^+$ ratio
for the $^{96}$Ru+$^{96}$Ru reaction at 0.4 AGeV.}\label{rpi-nz}
\vspace{-0.3cm}
\end{figure}

Shown in Fig.\ \ref{rpi-nz} are the calculated $\pi^-/\pi^+$ ratios in comparison
with the FOPI data at 0.4 AGeV with the reduced impact parameter $b_0\equiv b/b_{max}\leq 0.15$ as a
function of the neutron/proton ratio of the reaction system. The
inset shows the $\pi^-/\pi^+$ ratio as a function of $b_0$ for the
$^{96}$Ru+$^{96}$Ru reaction at 0.4 AGeV. It is seen that both
the data and the calculations exhibit very weak $b_0$ dependence for
the $\pi^-/\pi^+$ ratio, even for mid-central reactions where we
found that the multiplicities of both $\pi^-$ and $\pi^+$ vary
appreciably with the $b_0$. For the
symmetric $^{40}$Ca+$^{40}$Ca and the slightly asymmetric
$^{96}$Ru+$^{96}$Ru reactions, calculations using both $x=1$ and
$x=0$ can well reproduce the FOPI data. Interestingly, for the more
neutron-rich reactions of $^{96}$Zr+$^{96}$Zr and
$^{197}$Au+$^{197}$Au calculations with $x=0, 0.5$ and $1$ are
clearly separated from each other. The FOPI data favors clearly the
calculation with $x=1$. As shown in Fig. \ref{fig:esym}, with $x=1$
the \esym at $\rho\geq 2\rho_0$ reached in the reaction is very
small, leading to a rather high $N/Z$ in the
participant region and thus the larger $\pi^-/\pi^+$ ratio observed.
We note here that in the present study we used directly the
condition $b_0\leq 0.15$ in selecting the most central events. On
the other hand, in the data analyses and the IQMD calculations, the
distribution of ERAT (ratio of transverse to longitudinal kinetic
energies) was used in determining the centrality of the reaction.
Nevertheless, given the fact that the $\pi^-/\pi^+$ ratio is almost
a constant within error bars over a large range of $b_0$, more
elaborate selection of the most central events by matching the
calculated and the experimental ERAT distributions may change
somewhat quantitatively the $\pi^-/\pi^+$ ratio, but it is not
expected to change qualitatively our conclusions. For comparisons,
the IQMD result from ref.\cite{Rei07} is also shown. As seen in Fig.
\ref{fig:esym}, the \esym used in the IQMD, the MDI \esym with x=0
and the APR prediction are all very close to each other for $\rho_0
<\rho \leq 3\rho_0$. Thus, not surprisingly, the $\pi^-/\pi^+$
ratios from the IQMD and the IBUU04 with x=0 are very close too.
Moreover, they both grow approximately according to the scaling
$\pi^-/\pi^+\approx (N/Z)^2$ predicted by the $\Delta(1232)$
resonance model but fall far below the FOPI data. Our calculations
with varying values of $x$ indicate that a strong symmetry potential
is at work at this energy as one expects. While not perfectly
reproduced by our calculations even with $x=1$, the FOPI data
suggest unambiguously that the \esym is rather soft at
supra-saturation densities compared to the APR prediction.

\begin{figure}[h]
\vspace{-0.2cm}
\centering\includegraphics[width=0.7\columnwidth,scale=0.4]{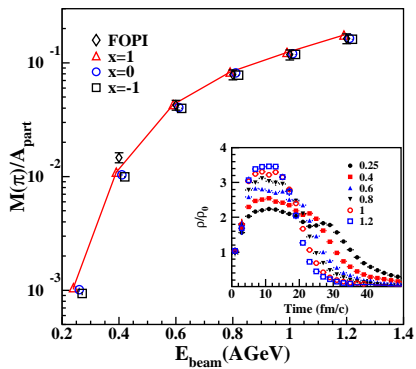}
\vspace{-0.4cm} \caption{(Color online)Excitation function of the
pion multiplicity per participant $M_{\pi}/A_{part}$ in the most
central Au+Au collisions. The inset shows the time evolution of
the central density.} \label{fig:multi} \vspace{-0.2cm}
\end{figure}

Since the Au+Au collision is most sensitive to the \esym among the
reactions considered here, we now turn to the excitation functions
of the pion yield and the $\pi^-/\pi^+$ ratio for the most central
Au+Au reactions. Both the pion yield and the $\pi^-/\pi^+$ ratio do
not change noticeably within error bars with the fine sub-division
of the impact parameter within $b_0\leq 0.15$. We thus compare
simply calculations for head-on collisions with the FOPI data for
the most central reactions. Figure \ref{fig:multi} displays the
excitation function of the pion multiplicity per participant
$M_{\pi}/A_{part}$. In order to compare with the data directly, the
total pion multiplicity is obtained from the charged pions only by
using $1.5\times(M_{\pi^-}+M_{\pi^+})$ and the number of
participants is calculated from $0.9\times A_{sys}$ where $A_{sys}$
is the total mass of the colliding system as done in the data
analysis\cite{Rei07}. It is seen that the results of the
calculations are in reasonably good agreement with the available
data. We notice here that, unlike the $\pi^-/\pi^+$ ratio which has
the advantage of reducing significantly not only systematic errors
but also effects of isoscalar nuclear potentials, the pion yield
also depends appreciably on the EOS of symmetric nuclear
matter\cite{Rei07,Sto86}. The inset illustrates the time evolution
of the central density with the maximum value varying from about
$2.2\rho_0$ to $3.5\rho_0$ for the beam energy from 0.25 to 1.2
AGeV.

\begin{figure}[h]
\vspace{-0.5 cm}
\centering\includegraphics[width=0.8\columnwidth,scale=0.4]{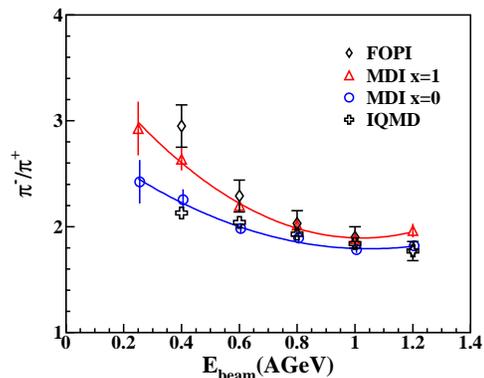}
\vspace{-0.5cm} \caption{(Color online)Excitation function of the
\rpi ratio in central Au+Au collisions calculated with the IBUU04 in
comparison with the FOPI data and the IQMD prediction.}
\label{fig:ratio}
\end{figure}

Shown in Fig.\ \ref{fig:ratio} are the excitation functions of the
$\pi^-/\pi^+$ ratio calculated with the IBUU04 and the IQMD in
comparison with the FOPI data. First of all, the decreasing trend of
the \rpi ratio with the increasing beam energy is well reproduced by
all calculations. Most interestingly, IBUU04 calculations with $x=1$
can best describe the FOPI data over the whole energy range.
Moreover, the \rpi ratio is more sensitive to the \esym at lower beam
energies as expected.

Putting together all available information we infer that the \esym
reaches a maximum somewhere between $\rho_0$ and $2\rho_0$ before it
starts decreasing at higher densities. This indicates the importance
of the development at NSCL/MSU~\cite{Abi07}
and RIKEN~\cite{Mur08} of Time Projection Chambers (TPC) to measure
\rpi ratios in heavy-ion reactions induced by radioactive beams from
about 120 to 400 AMeV. These measurements around the $E^{\pi}_{th}$
can help constrain the \esym at densities from about $\rho_0$ to $2\rho_0$.
Moreover, the neutron-proton differential flow measurements planned at
SIS/GSI\cite{Roy08} and possibly also at the future FAIR/GSI will
not only test our findings here but may also enable the exploration
of the \esym at even higher densities. Generally speaking, there are
limitations on how much one can learn reliably about the high
density \esym from high energy heavy-ion collisions. For instance,
to reach higher baryon densities by using more energetic beams, more
baryon and meson resonances are excited. Moreover, with the
increase of the beam energy, the collision becomes more transparent~\cite{Fu08}.
Therefore, the hot and dense system formed is further away from the
pure nucleonic matter in thermodynamical equilibrium.
It thus becomes more difficult to model and relate properties of
this system with the \esym of nucleonic matter. In this regard,
neutron stars are unique laboratories for probing the \esym of cold
high density nucleonic matter. On the other hand, astrophysical studies of
neutron stars have their own limitations and challenges. It is thus important
to do cross checking between the two fields. If our conclusion on the \esym is
confirmed by more experimental and theoretical studies, it not only
posts a serious challenge to some nuclear many-body theories but
also has important implications on several critical issues in
nuclear astrophysics, such as, the cooling of proto-neutron stars,
the possible formation of polarons due to the isospin separation
instability\cite{Kut94,Szm06}, the possible formation of quark
droplets\cite{Kut00} and hyperons\cite{Ban00} in the core of neutron
stars.

In summary, circumstantial evidence for a rather
soft nuclear symmetry energy \esym at $\rho\geq 2\rho_0$ compared to
the APR prediction was obtained from analyzing the recent FOPI data
on the \rpi ratio using the IBUU04 transport model. Further
experimental and theoretical confirmations of our findings are
important.

This work was supported in part by the National Natural Science
Foundation of China under grants 10675148, 10575071 and 10675082,
MOE of China under project NCET-05-0392, Shanghai Rising-Star
Program under Grant No.\ 06QA14024, the SRF for ROCS, SEM of
China, and the National Basic Research Program of China (973
Program) under Contract No. 2007CB815004, the US National Science
Foundation Awards PHY-0652548 and PHY-0757839, the Research
Corporation under Award No. 7123 and the Texas Coordinating Board
of Higher Education Award No. 003565-0004-2007.

\end{document}